\documentclass[prl,twocolumn,showpacs,floatfix,superscriptaddress]{revtex4}
\usepackage{amsmath,amsfonts}
\usepackage{graphicx}
\usepackage{color}
\usepackage{enumerate}
\usepackage{bbm}
\usepackage{wrapfig}
\usepackage{bm}
\usepackage{hyperref}
\usepackage{float}

\newcommand{\bk}{\mathbf{k}}

\newcommand{\veps}{\varepsilon}

\newcommand{\beq}{\begin{equation}} 
\newcommand{\eeq}{\end{equation}}

\newcommand{\nn}{\nonumber}

\begin{document}



\title{Unveiling odd-frequency pairing around a magnetic impurity in a superconductor}

\author{Vivien Perrin}
\affiliation{Universit\'e Paris-Saclay, CNRS, Laboratoire de Physique des Solides, 91405, Orsay, France}

\author{Fl\'avio L. N. Santos}
\affiliation{Universit\'e Paris-Saclay, CNRS, Laboratoire de Physique des Solides, 91405, Orsay, France}
\affiliation{Departamento de Fisica, Universidade Federal de Minas Gerais,
Caixa Postal 702, Belo Horizonte, MG, 30123-970, Brazil}

\author{Gerbold C. M\'enard}
\affiliation{Institut des NanoSciences de Paris, Sorbonne Universit\'e and CNRS-UMR 7588, 75005 Paris, France}
\author{Christophe Brun}
\affiliation{Institut des NanoSciences de Paris, Sorbonne Universit\'e and CNRS-UMR 7588, 75005 Paris, France}
\author{Tristan Cren}
\affiliation{Institut des NanoSciences de Paris, Sorbonne Universit\'e and CNRS-UMR 7588, 75005 Paris, France}
\author{Marcello Civelli}
\affiliation{Universit\'e Paris-Saclay, CNRS, Laboratoire de Physique des Solides, 91405, Orsay, France}
\author{Pascal Simon}
\affiliation{Universit\'e Paris-Saclay, CNRS, Laboratoire de Physique des Solides, 91405, Orsay, France}

\date{\today}





\begin{abstract}
We study the unconventional superconducting correlations caused by a  single isolated magnetic impurity in a conventional s-wave superconductor. Due to the local breaking of 
time-reversal symmetry, the impurity induces unconventional superconductivity which is even in both space and spin variables but odd under time inversion.
We  derive an exact  proportionality relation between the even-frequency component of the local electron density of states and the imaginary part of the odd-frequency local pairing function.
By applying this relation to scanning tunneling microscopy spectra taken on top of magnetic impurities immersed in a Pb/Si(111) monolayer, we show experimental evidence of the occurrence of the odd-frequency pairing in these systems and explicitly extract its superconducting function from the data.
\end{abstract}


\maketitle


{\em Introduction.}
Due to Fermi-Dirac statistics, the two-electron pairing correlation function at different times
$t_1$ and $t_2$ has to be anti-symmetric under exchange of the two electrons or equivalently under the exchange of all their labels. These include time, spin, position and possibly other orbital degrees of freedom. In the conventional s-wave superconductor, the pairing function corresponds to an equal time, s-wave and spin-singlet pairing (assuming a single band) while the coveted p-wave superconductor corresponds to an equal time, p-wave, spin-triplet pairing function \cite{Sigrist1991}. In the former case the sign change of the pair amplitude is provided by the spin variable, while in the latter by the space one. 
However there is a possibility that the sign
may change under exchange of the two different time coordinates $t_1\ne t_2$.
More than four decades ago, Berezinskii proposed this possibility, the odd-frequency (odd-$\omega$) 
pairing (thus odd under time exchange) in the s-wave triplet pairing of He$^3$
 \cite{Berezinski} (see \cite{Tanaka2012,Linder2017} for recent reviews of this long debated field). It was subsequently considered that odd-$\omega$ pairing can also be intrinsically generated in superconductors \cite{Belitz1991,Balatsky1992,Abrahams1995}, or  in heavy fermions compounds described by a Kondo lattice model \cite{Coleman1993,Coleman1994, Coleman1997}.

The field boomed when Bergeret {\it et al.} realized that odd-$\omega$ pairing should appear in heterostructures made of a conventional s-wave superconductor and a ferromagnet \cite{Bergeret2001,Bergeret2005}. Such a platform has the key advantage of realizing odd-$\omega$ pairing in a controllable fashion with well understood materials. Conversely to the previous studies, the 
odd-$\omega$ pairing in superconductor/ferromagnet hybrid structures is the result of a proximity effect where the ferromagnet induces a
spin-singlet to spin-triplet conversion of Cooper pairs. Such conversion actually allows Cooper pairs to propagate robustly far away in the ferromagnet \cite{Quay2012}. This opens the exciting possibility to achieve spintronics with superconductors \cite{Eschrig2015,Linder2015}.
 Subsequent studies demonstrated that odd-frequency pairing in fact appears in a wide variety of physical systems as a result of symmetry breaking. For example, odd-$\omega$ pairing can be realized in non-magnetic junctions due to spatial parity breaking at the interface \cite{Tanaka2007a,Tanaka2007b,Eschrig2007},
which allows the conversion from s-wave to p-wave orbital symmetry.
According to these predictions, odd-$\omega$ pairing should be rather ubiquitous in hybrid systems. This also explains the recently established connection between odd-$\omega$ pairing and the physics of Majorana fermions \cite{Tanaka2013} (see \cite{Linder2017,Cayao2020} for recent reviews). However, there is not yet any clear and direct experimental evidence
of odd-$\omega$ superconductivity, though  spectroscopic
signatures in the density of states were reported in Nb superconducting films coupled to epitaxial Ho by proximity effect \cite{Robinson2015}.

Here we show evidence of the existence of odd-$\omega$ pairing in the simplest hybrid system: a single magnetic impurity immersed in a conventional s-wave, spin singlet, even-$\omega$ superconductor. First, our analysis shows that on the magnetic impurity site a s-wave (local), spin triplet and  odd-$\omega$ superconducting component arises from the breaking of the rotational symmetry. We then establish 
an exact proportionality relation between the even-$\omega$ component of the local-impurity electron density of states (LDOS) 
and the imaginary part of the odd-$\omega$ superconducting function and provide expressions for the proportionality coefficients, 
which only depend on the parameters characterizing the magnetic impurity.
We finally apply these results to account for the local density of states measured with scannning tunelling spectroscopy (STS) on top of magnetic impurities immersed in a superconducting monolayer of Pb/Si(111). 
This provides the experimental
evidence of the presence of the odd-$\omega$ pairing component. Moreover, we are able to extract and explicitly display 
the superconducting odd-$\omega$ pairing function.

{\em Local pairing functions.}
Due to the fermionic anti-commutation relations, the retarded and advanced superconducting functions 
$F^{R}_{\alpha,\beta}= -i\theta(t-t')\langle\lbrace c_\alpha(t),c_\beta(t')\rbrace\rangle$, $F^{A}_{\alpha,\beta}= i\theta(t'-t)\langle\lbrace c_\alpha(t),c_\beta(t')\rbrace\rangle$
are related, 
$F^R_{\alpha,\beta}(t,t')= -F^A_{\beta,\alpha}(t',t)$, or in frequency space 
$F^R_{\alpha,\beta}(\omega)= -F^A_{\beta,\alpha}(-\omega)$, under particle exchange. 
Here the symbol $\langle...\rangle$ is a short-hand notation denoting thermal average with respect to equilibrium state.
$\alpha,\beta= \, (\vec{r}, \uparrow, \downarrow...)$ are a priori any relevant set of quantum numbers, depending on the system. 
As here we study the local impurity ($\vec{r}=0$), only spin variables are considered $\alpha,\beta= \, (\uparrow, \downarrow)$. 
We can always choose  the  order-parameter of the bare BCS superconductor to be real. Therefore, it follows
that $F^R(\omega)=(F^A(\omega))^*$.

Thus, there are only two possible ways for the local retarded pairing functions to satisfy these relations:
\begin{align}
   1. &F^R_{\uparrow,\downarrow}(\omega)= F^{R*}_{\uparrow,\downarrow}(-\omega) &{\rm Even}~\omega~;~ {\rm spin~ singlet},\\
    2. &F^R_{\uparrow,\downarrow}(\omega)= -F^{R*}_{\uparrow,\downarrow}(-\omega) &{\rm Odd}~ \omega~;~{\rm spin~ triplet}.
\end{align}
It is therefore convenient to decompose $F^{R}_{\uparrow,\downarrow}(\omega)$ in even$-\omega$ (spin-singlet) and an odd$-\omega$ 
(spin-triplet) components:
\begin{align}
 F^R_{even/odd}(\omega) = \frac{1}{2}[F^R_{\uparrow,\downarrow}(\omega)\pm F^{R*}_{\uparrow,\downarrow}(-\omega)].
 \end{align}
This implies that $\Re{F^R_{even}}(\omega)$, $\Im{F^R_{odd}}(\omega)$ are even functions while $\Im{F^R_{even}}(\omega)$, $\Re{F^R_{odd}}(\omega)$ are odd functions of frequency (here $\Re,\Im$ correspond to the real and imaginary part respectively).
A $F^R_{odd}(\omega)$ component in the total superconducting function is therefore the fingerprint of odd-$\omega$ superconductivity. The difficulty in proving the existence of odd-$\omega$ superconductivity relies then 
in extracting the superconducting function from spectral quantities. Our goal is to show that $F^R_{odd}(\omega)$ can be indeed
 extracted from the LDOS measured  with STS on a magnetic impurity site.

{\em Model Hamiltonian and Dyson equation.}
We consider a single magnetic impurity in a s-wave homogeneous superconductor.
We use the well-known Yu-Shiba-Rusinov (YSR) model \cite{Yu1965,Shiba1968,Rusinov1969} to describe this hybrid system (see \cite{Balatsky2006} for a review). 
The Bogoliubov-de Gennes (BdG) Hamiltonian reads

\begin{align}
\mathcal{H}&=\int d\textbf{r}
\,\Psi^\dagger(\textbf{r})
\begin{pmatrix}
\veps(-i\mathbf{\nabla}_{\textbf{r}})&\Delta(\textbf{r})\\ 
\Delta(\textbf{r})&-\veps(-i\mathbf{\nabla}_{\textbf{r}})
\end{pmatrix}\Psi(\textbf{r})\nn\\
&+
\Psi^\dagger(\textbf{r}=\textbf{0})
\begin{pmatrix}
V-J&0\\0&-(V+J)
\end{pmatrix}\Psi(\textbf{r}=\textbf{0}),
\label{eq:ham}
 \end{align}
where  $\Psi^{T}(\textbf{r})=(c_{\uparrow}(\textbf{r}),c^\dagger_{\downarrow}(\textbf{r}) )$ is a 2-component Nambu spinor. The superconductor is characterized by the metallic dispersion relation $\veps(\bk)$ and a real pairing potential $\Delta(\textbf{r})$. The magnetic impurity is modeled as a classical exchange field of strength $J$ and a potential scattering $V$. We neglect any momentum dependence of these local couplings as this does not play any role here. Due to the presence of  impurities, the pairing gap may be weakly affected around the impurity\cite{Schrieffer1997,Meng2015}. We neglect in what follows any spatial dependence of $\Delta(\textbf{r})$ as we did not observe such gap renormalization in our experimental data. However, because our predictions do depend only on the {\em local} density of states at the position of the impurity, only 
$\Delta(\textbf{r}=\textbf{0})$ matters and our theory can therefore incorporate a renormalization of the gap.

 We use the Nambu-Gorkov Green's function to completely describe
 the local one-particle electronic properties of the system
\begin{align}\label{eq:nambu}
    \hat{G}^R(t,t')=& -i\theta(t-t')\langle\Psi(\textbf{0},t)\Psi^\dagger(\textbf{0},t')\rangle\nonumber \\
		=&
    \begin{bmatrix}G^R_\uparrow(t-t')&F^R_{\uparrow,\downarrow}(t-t')\\-F^R_{\downarrow,\uparrow}(t-t')^*&-G^R_{\downarrow}(t-t')^*\end{bmatrix},
\end{align}
where $\hat{G}^R(t,t')= -i\theta(t-t')\langle \lbrace \Psi^{}(t), \Psi^{\dagger}(t') \rbrace\rangle$ contains both the normal 
$G^R$ (diagonal) and anomalous $F^R$ (off-diagonal) components.

\begin{figure}[t]
        \centering
   \includegraphics[width=0.7\linewidth]{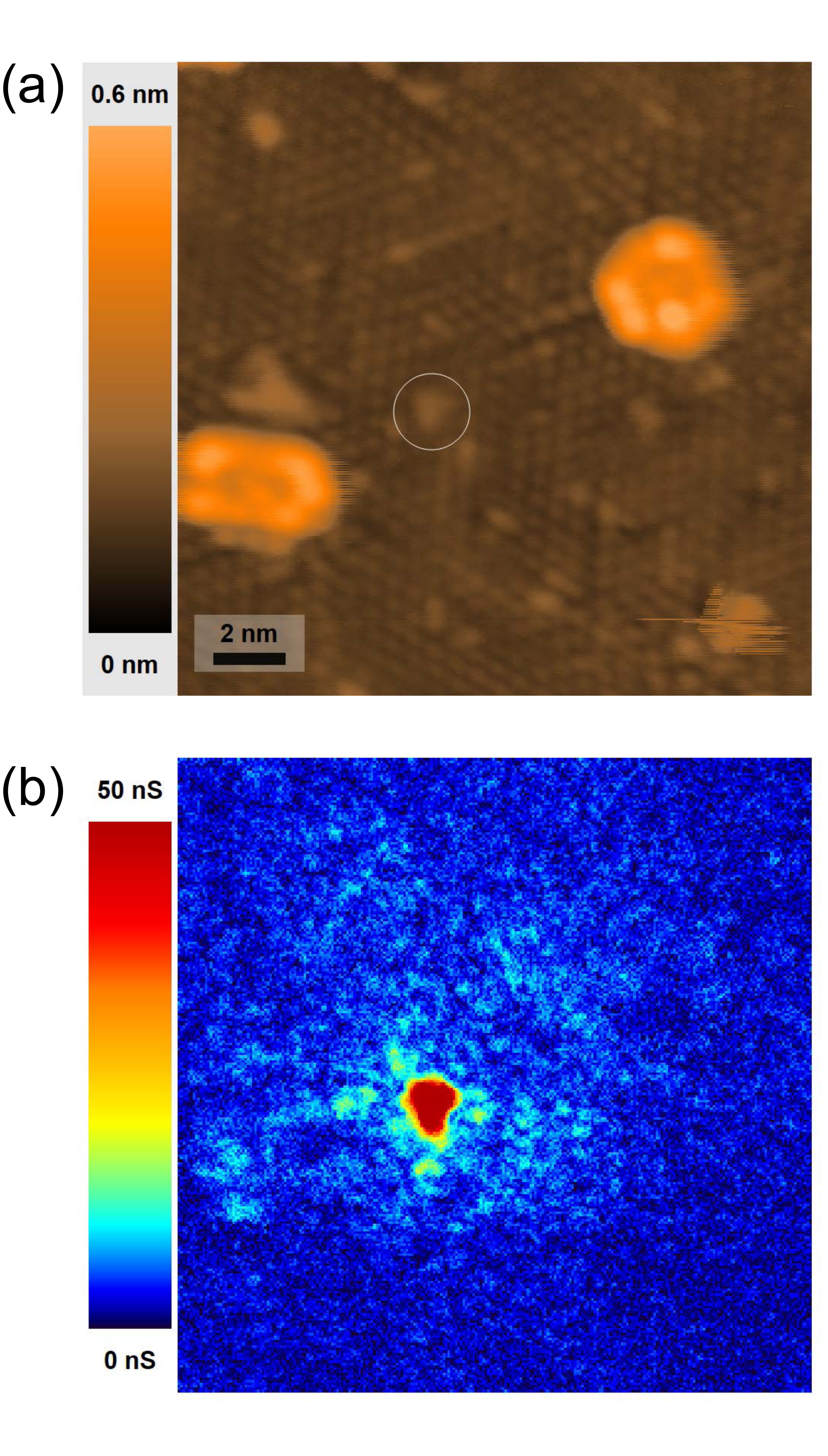}
\caption{(a): scanning tunnelling microscopy image of the Pb monolayer where a magnetic defect is present as a triangular protrusion. (b): corresponding conductance map measured at the Fermi level by scanning tunnelling spectroscopy at 320mK.}
 \label{MAP}
\end{figure}
The full  Nambu-Gorkov local Green's function in Fourier space at the position of the impurity can be computed 
using the Dyson equation 
\begin{align}
 \hat{G}^R(\omega)=\hat{g}^R(\omega)+\hat{g}^R(\omega)\hat\Sigma\hat{G}^R(\omega),\end{align}
where $\hat{g}^R(\omega)$ is the local Green's function of the  bulk superconductor in the absence of the magnetic impurity (\textit{i.e.} the bare one) and 
\begin{align}
   \hat{\Sigma}=\begin{bmatrix}V-J-i\Gamma&0\\0&-(V+J)-i\Gamma\end{bmatrix},
\end{align}
is the local self-energy in Nambu-space at the position of the impurity. Note that we also include a phenomenological Dynes broadening $\Gamma$ to this self-energy. 
The Dyson equation is easily solved as \cite{sm}
\beq
\hat{G}^R(\omega)=[\hat{g}^R(\omega)^{-1}-\hat{\Sigma}]^{-1}.\eeq
 Note the only assumptions we made at this point are that there is s-wave pairing and that the spin is locally a good quantum number.
For $\vert\omega\vert<\Delta$ inside the superconducting gap, the bare Green's function is real in our gauge choice.  Therefore the imaginary part of $\hat{G}^R(\omega)$ is a sum of Dirac-like distributions located at the poles of $1/{\rm Det}[\textbf{1}-\hat{g}^R(\omega)\hat{\Sigma}]$. 
These are the  YSR in-gap 
spin-polarized bound states \cite{Yu1965,Shiba1968,Rusinov1969,Balatsky2006}.

{\em Magnetic impurities in a Pb/Si(111) substrate.}
%
Due to recent enormous progress in the energy and spatial resolution of STS, YSR  states are now very well characterized experimentally (see \cite{Heinrich2018} for a recent review). This revival of the physics of the YSR states has  also been motivated by the study of chains
of magnetic atoms on a  superconducting substrate which has attracted a considerable attention in the past years \cite{Choy2011,Nakosai2013,NP2013,Braunecker2013,Klinovaja2013,Vazifeh2013,Pientka2013,Pientka2014,Poyhonen2014,Reis2014,Kim2014,Li2014,Heimes2014,Brydon2015,Ojanen2015a,Peng2015,Rontynen2015,Hui2015,Braunecker2015,Li2016a,Rontynen2016}. Recent experiments on such systems have revealed 
the existence of zero bias  peaks spatially localized on the ends of the chains which have been interpreted as signatures of Majorana bound states \cite{NP2014,Pawlak2016,Ruby2015,Yazdani2017,Ruby2017,Kim2018}. \\
Here we consider STS data of magnetic impurities immersed in a Pb/Si(111) monolayer (see Figs 1 and 2). The Pb monolayer corresponds to a nominal  coverage of 4/3 with the stripe incommensurate reconstruction. This Pb monolayer was 
shown to be superconducting below 1.8K \cite{Zhang2010}. This system does not show any in-gap states in the presence of a strong 
non-magnetic disorder as expected for a s-wave superconductor \cite{Brun2014}. However, in presence of magnetic defects, YSR states 
manifest themselves by huge pairs of conductance peaks in a well-defined gap \cite{Menard2019}.
Fig. 1a shows a scanning tunnelling microscopy image of the Pb monolayer where a magnetic defect is present as a triangular protrusion. The corresponding conductance map measured at the Fermi level by STS at 320mK is shown in Fig. 1b. One can see a red spot on top of the defect that corresponds to a very strong YSR state, that is surrounded by a speckles like pattern due to the decaying YSR wave function scattered by the atomic disorder of the monolayer.

\begin{figure}[t]
        
        \includegraphics[width=0.75\linewidth]{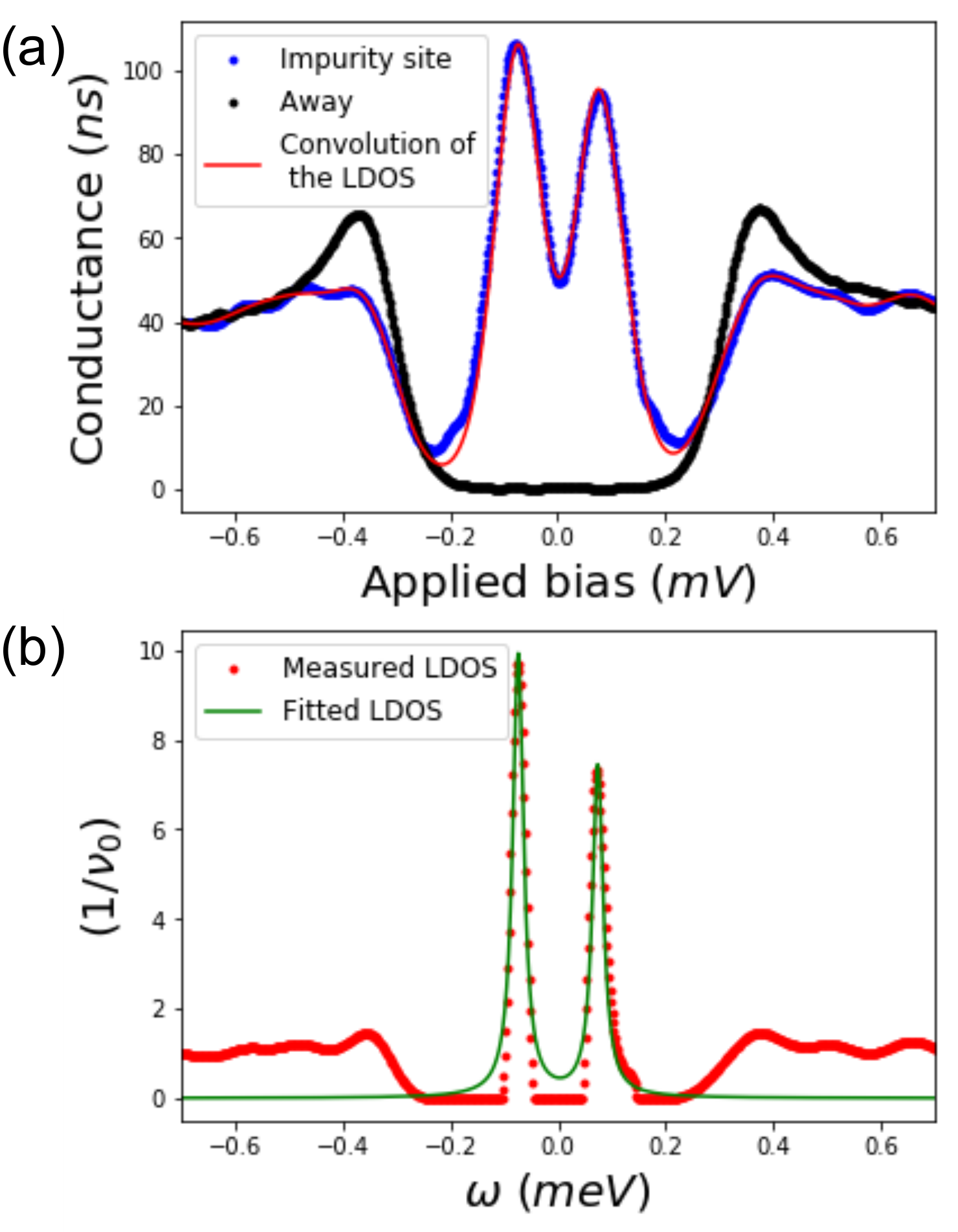}
\caption{(a): Measured differential conductance at the impurity site (blue dots), far away from the impurity (black dots) and the one obtained from convolution of the extracted LDOS with the Fermi-Dirac derivative (red line). (b):  LDOS obtained after deconvolution taking into account the finite temperature T=320mK (red dots). LDOS fitted with Eq. (\ref{eq:fit}) by using the parameter set: $E_0=0.074meV$, $\eta=0.012meV$, $u^2=0.28\nu_0$, $v^2=0.38\nu_0$ (green curve).}
 \label{DOS1}
\end{figure}

Fig. \ref{DOS1}a shows a spectrum (black curve) taken far from the impurity that corresponds to a BCS gap of 0.38 meV with a Dynes broadening $ \Gamma$=0.004 meV convoluted by a thermal broadening due to the finite temperature of 320 mK. By contrast a spectrum taken on top of the impurity (blue dots) exhibits a strong pair of YSR peaks in the gap.
Moreover, we always observe a 
single pair of conductance peak meaning that only one YSR state is present or at least that the eventual multiplet is degenerated 
up to the experimental resolution. Note that as the Pb monolayer is a 2D superconductor, the YSR states extend very 
far from the impurities (typically tens of nanometers) \cite{Menard2015}. However, we focus here  on spectra taken on top of the impurities.

{\em Local density of states and odd-$\omega$ pairing.}
From the above experimental plots, we can safely assume that the YSR poles are well-separated in energy. We show then that approximate expressions of the YSR states can 
be obtained as a function of $J,V$ and of the expressions of the bare Green's functions 
and explicitly extract the odd-$\omega$ parts of the superconducting function.

The electronic LDOS  measured by the differential conductance in STS is defined by
$\rho(\omega)=-\frac{1}{\pi}\Im\left\{\hat{G}^R_{11}(\omega)+\hat{G}^R_{22}(-\omega)\right\}$.
After some algebra, the LDOS can be expressed as a linear combination of odd/even-frequency pairings as
\begin{align}
    \rho(\omega)\approx C_e(E_0)\times \Im{F^R_{odd}}(\omega) 
		+ C_o(E_0)\times \Im{F^R_{even}}(\omega),
\end{align}
where $E_0$ is the YSR bound state energy,
\begin{align}
     C_e(E_0)&=\frac{ 2J A(E_0)
		-g^R_\uparrow(E_0)+g^R_\downarrow(-E_0)}{\pi f^R_{\uparrow\downarrow}(E_0)},
		\\
     C_o(E_0)&=\frac{2V A(E_0)
		-g^R_\uparrow(E_0)-g^R_\downarrow(-E_0)}{\pi f^R_{\uparrow\downarrow}(E_0)},
\end{align}
and $A(E_0)=f^R_{\uparrow\downarrow}(E_0))^2+ g^R_\uparrow(E_0)g^R_\downarrow(-E_0)$.
In these expressions, $f$ and $g$ are the bare substrate Green's function defined according to \eqref{eq:nambu}. Note that  the broadening $\Gamma$ is assumed not to depend on the energy in this low-energy range.
Thus the even/odd-$\omega$ components of the LDOS defined as $\rho_{even/odd}(\omega)=(\rho(\omega)\pm \rho(-\omega))/2  $ are directly proportional to the imaginary part of odd/even-frequency pairing functions respectively:
\begin{align}
    \rho_{even/odd}(\omega)=C_{e/o}(E_0)\times \Im{F^R_{odd/even}}(\omega).
\end{align}
We have therefore derived  a general proportionality relation between the even-$\omega$ part of the LDOS and the imaginary part of the odd-$\omega$ anomalous pairing function. This relation has the strong physical implication that as soon as there exists some in-gap YSR state, there is a local odd-$\omega$ pairing around the impurity site. Note that this is a direct consequence of the magnetic impurity locally breaking time-reversal symmetry.

The proportionality coefficient does in general depend on the way the substrate is modeled. However, in most physically relevant cases the Fermi energy of the substrate is the largest energy scale and  the normal DOS can be approximated by its value at the Fermi Energy, $\nu_0$. 
Considering that we have a single YSR in-gap bound state, its energy $E_0$ is given by
\begin{equation}
    E_0=\Delta\frac{1-\alpha^2+\beta^2}{\sqrt{(1-\alpha^2+\beta^2)^2+4\alpha^2}},
\end{equation}
where $\alpha=\pi\nu_0 J$ and $\beta=\pi\nu_0 V$ \cite{Rusinov1969}.
This allows one to easily express the proportionality coefficient,
\begin{align}
    C_e(E_0)&=-\frac{2}{\Delta}[E_0+\pi J \nu_0\sqrt{\Delta^2-E_0^2}]\nn\\
	&=-\frac{2}{\pi}\frac{1+\beta^2+\alpha^2}{\sqrt{(1-\alpha^2+\beta^2)^2+4\alpha^2}}.	
\end{align}
It is important to emphasize here that, contrary to $C_o(E_0)$ which vanishes in the case of a pure magnetic impurity ($V=0$), $C_e(E_0)$ never vanishes. 
We can thus always evaluate $\Im{F^R_{odd}}(\omega)=\rho_{even}(\omega)/C_e(E_0)$. Notice that  $\pi|C_e^{-1}|/2\in ]0,1] $. Given $\rho_{even}(\omega)$, the  odd-$\omega$ pairing function 
is maximal for $|C_e^{-1}|=2/\pi$ which is reached for $\beta=0$. This corresponds to a pure magnetic impurity. The opposite limit $|C_e^{-1}|\to 0$ is reached only for $\beta^2=\alpha^2+1=\infty$ which are unphysical values. Even for extremely large values of $\alpha,\beta\sim 4$, we can still obtain a lower bound for $|\Im{F^R_{odd}}(\omega)|$ \cite{sm}. Inside the gap, the pairing  thus has the same order of magnitude as the LDOS.

{\em Protocol to extract odd$-\omega$ pairing.} Let us now provide an efficient protocol to extract the imaginary part of the local odd frequency pairing function $\Im{F^R_{odd}}(\omega)$ around the impurity from LDOS spectroscopic measurements  performed in the tunneling regime. The differential conductance spectrum $dI/dV$ measured locally corresponds to the convolution of the local density of state $\rho(\omega)$ with the derivative of the Fermi-Dirac distribution at the experimental temperature. Once $\rho(\omega)$ is measured,
we normalize it in units of the normal-state DOS at the Fermi-level denoted $\nu_0$.
In order to extract a reliable estimate of  $\Im{F^R_{odd}}(\omega)$ we perform some simple data analysis.
To do so, we follow \cite{Ruby2015b} and  assume that the YSR is well approximated by the following retarded Green's function

\begin{align}
    \label{GFapprox}
    \hat{G}(\omega)&=\frac{1}{\omega+i\eta-E_0} \, \begin{bmatrix}u^2&uv\\uv&v^2\end{bmatrix},\\
    u^2,v^2&=2\pi\alpha\nu_0\Delta\frac{1+(\alpha\pm \beta)^2}{((1-\alpha^2+\beta^2)^2+4\alpha^2)^{3/2}},
\end{align} 
where, $u$ and $v$ are the electron and hole 
components of the YSR state.
This expression is obtained from the exact solution of the Dyson equation after a first-order expansion around  energy $E_0$ \cite{Ruby2015b}. The phenomenological parameter $\eta$ introduced here takes into account the broadening of the YSR peaks due to relaxation. It can be related to $\Gamma$ up to some renormalization coefficient. 
With this expression of the Green's function, one then obtains
\beq
\rho(\omega)=\frac{\eta u^2/\pi}{(\omega-E_0)^2+\eta^2}+\frac{\eta v^2/\pi}{(\omega+E_0)^2+\eta^2},\label{eq:fit}
\eeq
that we use to fit the experimental data and extract the parameters $u^2$, $v^2$ and the inverse lifetime $\eta$. With these values at hand, one obtains the coefficient $  C_e(E_0)= -\frac{u^2+v^2}{\pi uv}$ and thus 
$\Im{F^R_{odd}}(\omega)=\rho_{even}(\omega)/C_e(E_0)$. Note that $\Re{F^R_{odd}}(\omega)$ can be obtained by the Kramers-Kronig relation.

\begin{figure}[t]
        \centering
				\includegraphics[width=0.75\linewidth]{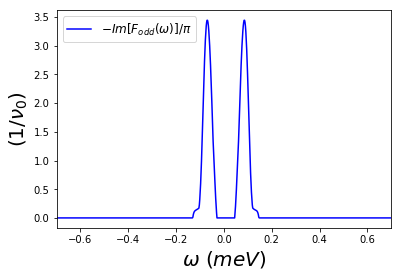}
\caption{$-Im{F^R_{odd}}(\omega)/\pi$ extacted at the position of the impurity (the values are relevant only inside the gap) from the measured LDOS shown in Fig. \ref{DOS1} }
 \label{ImF}
\end{figure}
{\em Application to magnetic impurities in a Pb/Si(111) substrate.}
We apply the previous protocol to extract $\Im{F^R_{odd}}(\omega)$ from the differential conductance 
on a magnetic impurity in a Pb/Si(111) substrate displayed in Fig. \ref{DOS1}b.
We extract the LDOS, and check that, once reconvoluted, it matches perfectly the original spectrum. We then fit it according to
Eq. (\ref{eq:fit}).
The fitted results for $\rho(\omega)$ 
show a good agreement with the data.  In order not to overestimate $\Im{F^R_{odd}}(\omega)$, we always take into account the parameters set which maximizes the ratio $v^2/u^2$. With the parameters $u,v,\eta$ at hand, we have thus access to $C_e(E_0)$.
We are finally able to obtain explicitely $-\Im{F^R_{odd}}(\omega)/\pi$ and display it in Fig. \ref{ImF}. 
Notice that it is symmetric in $\omega$ and its amplitude is comparable to $\rho(\omega)$.
We have applied this procedure to other sets of YSR states, fully confirming the above results \cite{sm}.

{\em Conclusion.} We show that an isolated magnetic impurity in a s-wave superconductor generates local pairing correlations which are odd in frequency. We provided a protocol to extract these anomalous pairing functions from STS measurements and apply it to 
data taken from a Pb/Si(111) monolayer   with magnetic impurities. Our theoretical/experimental analysis finally proves the occurrence of odd-$\omega$ pairing in the simplest magnetic-superconductor hybrid system.
While finalizing our manuscript
we learned about the theoretical work of D. Kuzmanovski {\it et al.} \cite{Balatsky2019} which 
has partial overlap with the theory part of our study. Differently from our work however, their study mainly focus 
on the spatial and frequency dependence on the spin-resolved LDOS in relation with odd-$\omega$ pairing.

{\em Acknowledgment.} We would like to thank  M. C. Aguiar, M. Aprili,  F. Massee, A. Mesaros, E. Miranda, A. Palacio-Morales, M. Rozenberg for interesting discussions and collaborations on related subjects. This work was supported by CAPES-COFECUB-0899/2018 (F.L.N.S., M.C., P.S.).


\bibliography{Biblio_Shiba}
\end{document}